\documentclass[aps,twocolumn,showpacs,preprintnumbers,nofootinbib,superscriptaddress]{revtex4}
\usepackage{amsmath} \usepackage{graphicx} \usepackage{amsfonts}
\usepackage{array} \usepackage{amsthm} \usepackage{bm}
\usepackage{palatino} \usepackage{mathpazo} 

\usepackage[breaklinks]{hyperref}
\usepackage{color}

\newcommand{\nn}{\nonumber}
\newcommand{\be}{\begin{equation}}
\newcommand{\ee}{\end{equation}}
\newcommand{\ba}{\begin{eqnarray}}
\newcommand{\ea}{\end{eqnarray}}
\newcommand{\bal}{\begin{align}}
\newcommand{\eal}{\end{align}}

\newcommand{\e}{{\rm e}}
\newcommand{\dd}{{\rm d}}

\newcommand{\bb}{\bibitem}

\newcommand{\om}{\omega}
\newcommand{\al}{\alpha}

\newcommand{\bt}{\beta}

\newcommand{\ga}{\gamma}

\newcommand{\ep}{\epsilon}

\newcommand{\ta}{\theta}

\newcommand{\vp}{\varphi}

\newcommand{\bw}{\begin{widetext}}
\newcommand{\ew}{\end{widetext}}

\begin{document}
\title{Wormhole solutions sourced by fluids, I: Two-fluid charged sources}

\author{Mustapha Azreg-A\"{\i}nou}
\affiliation{Ba\c{s}kent University, Faculty of Engineering, Ba\u{g}l\i ca Campus, 06810 Ankara, Turkey}


\begin{abstract}
We briefly discuss some of the known and new properties of rotating geometries that are relevant to this work. We generalize the analytical method of superposition of fields, known for generating nonrotating solutions, and apply it to construct massless and massive rotating physical wormholes sourced by a source-free electromagnetic field and an exotic fluid both anisotropic. Their stress-energy tensors are presented in compact and general forms. For the massive rotating wormholes there exists a mass-charge constraint yielding almost no more dragging effects than ordinary stars. There are conical spirals through the throat along which no local negative energy densities are noticed for these rotating wormholes. This conclusion extends to nonrotating massive type I wormholes derived previously by the author that seem to be the first kind of nonrotating wormholes with this property. Based on the classification made in J. Cosmol. Astropart. Phys. 07 (2015) 037 [arXiv:1412.8282]: ``Type I wormholes have their radial pressure dying out faster, as one moves away from the throat, than any other component of the stress-energy and thus violate the least the local energy conditions. In type II (resp. III) the radial and transverse pressures are asymptotically proportional and die out faster (resp. slower) than the energy density".
\end{abstract}


\maketitle

\section{Introduction\label{secI}}

Detection of a negative energy density of whatsoever source has not been made so far. That does not seem to be possible in the near future. However, its indirect effects have been measured for both the Casimir effect~\cite{cas1,cas2} and squeezed vacuum states where vacuum fluctuations are suppressed giving rise to regions of alternating positive and negative energy~\cite{vac1,vac2}.

The discussion of exotic matter involves negative energy densities necessary for sustaining the so-called wormholes, which are hypothesized tunnels connecting regions of spacetime. These have always been subject of intense research area. Many nonrotating and rotating wormholes have been found to general relativity and generalized theories of gravity (the list of references is too long to mention all of them). In this work we determine new rotating wormholes counterparts of existing nonrotating ones and attach physical interpretations to their sources. We develop a new analytical method based on the superposition of fields each attached to an anisotropic rotating fluid. The method is known for, and was applied to, nonrotating solutions~\cite{tol,snk,ex,BK}. The generalization of the method to rotating solutions necessitates the introduction of different rotating or comoving bases.

In the following section we discuss the properties of the metric of a rotating star and in Sec.~\ref{wor} we discuss and extend the properties of its special form used for generating rotating wormholes. In Sec.~\ref{frst} we show that the rotating wormhole derived by Teo~\cite{teo} using the metric of Sec.~\ref{wor}, which was shown to be sourced by no fluid~\cite{W3}, could be sourced by two rotating fluids.

In Sec.~\ref{scnd} we superpose a source-free electromagnetic field to an exotic matter to generate an exact redshift-free rotating wormhole solution, which turns out to be \emph{a rotating counterpart} of the Bronnikov-Ellis wormhole~\cite{BE1,Ellis}. Based on their asymptotic behavior, a classification of nonrotating wormholes into three types has been made in Ref.~\cite{types}: Type I wormholes have their radial pressure dying out faster, as one moves away from the throat, than any other component of the stress-energy. In type II (resp. III) the radial and transverse pressures are asymptotically proportional and die out faster (resp. slower) than the energy density. For type I wormholes, the violation of the local energy conditions (LECs) occurs in a narrow region adjacent to the throat~\cite{types}, while for type III wormholes, the region may extend to spatial infinity. In Ref.~\cite{types} we described this behavior by telling that type I wormholes violate the least--when compared to other types of wormholes--the LECs; said otherwise, they are yield the minimum violation of the LECs than the other types of wormholes, and that type III wormholes violate them relatively the most; that is, they cause the most harm to the LECs. In Sec.~\ref{gen} we generalize the procedure to rotating wormholes with redshift effects and construct a rotating counterpart of a nonrotating type I wormhole derived in Ref.~\cite{types}.

In all derived rotating solutions we provide compact expressions for the components of the SETs of the two fluids as well as an expression for evaluating the angular velocity $\om$ of the rotating wormhole. This velocity is determined upon requiring the components of the SETs, which generally depend on $\om$, to reduce to their static values, which do not depend on $\om$, in the limit of no rotation.

In this work we provide a first example of a \emph{nonrotating} massive type I wormhole where no local negative energy densities are noticed on conical spiraling paths through the throat. This refutes the belief that rotation is the only way to reduce the effects of exotic matter.

In Sec.~\ref{LEC} we address the question pertaining to the NEC and WEC for the rotating wormholes derived here and for their nonrotating counterparts. We show that on a cone of equation $\ta =\text{constant}$ it is possible to find ways through the throat where no local negative energy densities are noticed. This conclusion extends only to the nonrotating type I wormhole derived in Ref.~\cite{types} but not to the Bronnikov-Ellis one. We conclude in Sec.~\ref{secc}.

\section{On the metric of a rotating star \label{sta}}

Using the required symmetry properties of a stationary and axisymmetric spacetime that is circular (particularly the existence of two commuting Killing vectors $\partial_t$ and $\partial_{\varphi}$), the standard metric for a rotating star in equilibrium may be brought to the following form in quasi-isotropic coordinates~\cite{rot1,rot2} (see~\cite{Eric} for more details):
\begin{equation}\label{g1}
    \dd s^2=N^2\dd t^2-A^2(\dd R^2+R^2\dd\ta^2)-D^2R^2\sin^2\ta(\dd\varphi-\om\dd t)^2.
\end{equation}
In quasi-isotropic coordinates the equality $g_{RR}=g_{\ta\ta}/R^2$ is justified by the fact that all two-dimensional metrics are related by a conformal factor. Here ($N^2,A^2,D^2,\om$) are positive functions depending on ($R,\ta$). Under the further assumption that the star rotates slowly, it retains its spherical symmetry without being flattened, and this results in $g_{\ta\ta}=g_{\varphi\varphi}/\sin^2\ta$, that is, in $A^2=D^2$~\cite{Eric}. So, the metric of slowly rotating stars reads
\begin{equation}\label{g2}
    \dd s^2=N^2\dd t^2-A^2\dd R^2-A^2R^2[\dd\ta^2+\sin^2\ta(\dd\varphi-\om\dd t)^2].
\end{equation}

The form~\eqref{g2} is not convenient for constructing wormhole or black hole solutions. Introducing a new radial coordinate $r$:
\begin{equation}\label{g23}
    R\equiv R(r),
\end{equation}
we bring it to the form
\begin{equation}\label{g3}
    \dd s^2=N^2\dd t^2-\e^{\mu}\dd r^2-r^2K^2[\dd\ta^2+\sin^2\ta(\dd\varphi-\om\dd t)^2],
\end{equation}
first derived in~\cite{Har}. Here we have set
\begin{equation*}
\Big(A\frac{\dd R}{\dd r}\Big)^2=\e^{\mu(r,\ta)}\quad\text{ and }\quad A^2R^2=r^2K^2(r,\ta).
\end{equation*}
Notice that, in order to satisfy the symmetry requirements, if $A$ depends only on the radial coordinate, so are the functions ($\mu,K$):
\begin{equation}\label{g4}
A\equiv A(r)\Rightarrow \mu\equiv\mu(r)\;\text{ and }\; K\equiv K(r).
\end{equation}

The metric~\eqref{g3} has the further property that $g_{\ta\ta}$ and $g_{\vp\vp}/\sin^2\ta=-r^2K^2$ are everywhere equal; it is sufficient that they are equal on the axis of rotation ($\ta=0$ or $\ta=\pi$) for the metric~\eqref{g3} not to have a conical singularity on it~\cite{Eric}.

Initially derived for slowly rotating stars~\cite{Har}, however, the metric~\eqref{g3} has been used as a standard form for the discussion and derivation of rotating wormholes~\cite{teo}. This implicitly assumes the absence of effects due to centrifugal forces which cause the ``surface" of the rotating solution to flatten.

It is worth mentioning that static~\cite{scw} and rotating~\cite{rcw1,rcw2} cylindrically symmetric wormholes are now an active topic of research. The metric~\eqref{g1} in quasi-isotropic coordinates is also useful for describing solutions endowed with cylindrical symmetry. In fact, if the metric coefficients ($N^2,A^2,D^2,\om$) are all one-variable functions depending on the new radial variable $u$, then by the coordinate transformation $R^2=u^2+z^2$ and $\ta=\arctan(z/u)$, where $z$ is a longitudinal coordinate, one brings~\eqref{g1} to the metric~(3) of Ref.~\cite{rcw2} in the gauge $A=C$ according to the notation of that reference.

\section{On the metric of a rotating wormhole \label{wor}}

In this paper a prime notation $f'(r,\ta,\cdots)$ denotes the partial derivative of $f$ with respect to (wrt) $r$, and derivation wrt to other variables is shown using the indexical notation, as in $f_{,\ta}\equiv \partial f/\partial \ta$.

The work done in Ref.~\cite{MT}, concerning the construction of nonrotating wormhole solutions, has suggested the introduction of the shape function~\cite{teo} $B(r,\ta)$
\begin{equation}
 1-\frac{B(r,\ta)}{r}\equiv \e^{-\mu(r,\ta)},
\end{equation}
in terms of which the metric~\eqref{g3} takes the form
\begin{equation}\label{w1}
    \dd s^2=N^2\dd t^2-\frac{\dd r^2}{1-\frac{B}{r}}-r^2K^2[\dd\ta^2+\sin^2\ta(\dd\varphi-\om\dd t)^2],
\end{equation}
where $N^2>0$ to ensure that the metric does not have horizons. We may choose $K>0$, in that case $K$ is a nondecreasing function of $r$ ($K'>0$). Now, to ensure that the metric is free of singularity on the axis of rotation ($\ta=0$ or $\ta=\pi$), further regularity conditions must be imposed. These conditions include the elementary flatness constraint (absence of conical singularity on the symmetry axis), which has been discussed in the previous section, the spacelike nature constraint of the axial Killing vector $\partial_{\varphi}$ in a neighborhood of the axis, and the existence of a Taylor series with only positive integer powers of the local Cartesian coordinates ($x,y,z$) of any metric function. This last constraint implies that the derivatives of ($N,B,K$) wrt $\ta$ have to vanish on the axis of rotation~\cite{teo,Lobo}. For instance, we can write $B_{,\ta}$ as
\begin{multline}\label{rc1}
B_{,\ta}=x_{,\ta}B_{,x}+y_{,\ta}B_{,y}+z_{,\ta}B_{,z}\\
=r\cos\ta\cos\vp B_{,x}+r\cos\ta\sin\vp B_{,y}-r\sin\ta B_{,z}\,,
\end{multline}
which reduces to
\begin{equation}\label{rc2}
 B_{,\ta}=r\cos\ta B_{,\bar{r}}--r\sin\ta B_{,z}\,,
\end{equation}
(with $\bar{r}=r\sin\ta$) if axisymmetry is imposed. To avoid a jump discontinuity in the value of $B_{,\ta}$ on the axis of symmetry as $z$ changes sign ($\cos\ta$ changes from $-1$ to $+1$), we take $B_{,\ta}=0$ there. As we shall see below, the scalar curvature depends on $B_{,\ta\ta}$ (and on other second derivatives wrt $\ta$), so the vanishing of the $\ta$ derivatives of ($N,B,K$) ensures that the scalar curvature does not diverge on the axis.

Asymptotic flatness requires
\begin{equation}\label{w1b}
\lim_{r\to\infty}N^2=\lim_{r\to\infty}K^2=1,\quad \lim_{r\to\infty}B/r=0.
\end{equation}

As in the nonrotating case, the surface of the throat is defined by
\begin{equation}\label{w2}
B(r_0,\ta_0)=r_0.
\end{equation}
This provides $r_0$ as a function of $\ta_0$; that is, for a given value of $\ta_0$ we solve~\eqref{w2} for $r_0$ and we keep the largest value.

In this work we only consider analytic functions $B(r,\ta)$ admitting Taylor series about the point ($r_0,\ta_0$). The curvature scalar $\mathcal{R}$ is given in Eq.~(92) of Ref.~\cite{Lobo}, the only factors of it that may diverge are~\cite{Lobo}
\begin{equation}\label{w3a}
\frac{B_{,\ta}}{r-B}\quad\text{ and }\quad \frac{B_{,\ta\ta}}{r-B}+\frac{3}{2}\Big(\frac{B_{,\ta}}{r-B}\Big)^2,
\end{equation}
and all the other terms of $\mathcal{R}$ are finite on the throat and elsewhere~\cite{Lobo}. The second factor in~\eqref{w3a} has been missed in Ref.~\cite{teo}. Thus, the curvature scalar associated with~\eqref{w1} is nonsingular everywhere provided the values of $B_{,\ta}$ and $B_{,\ta\ta}$ are zero on the throat:
\begin{equation}\label{w3}
B_{,\ta}|_{(r_0,\ta_0)}=0\quad\text{ and }\quad B_{,\ta\ta}|_{(r_0,\ta_0)}=0.
\end{equation}
These two constraints remove any divergence of $\mathcal{R}$ but they do not ensure a well-defined value of it on the throat, for the limit as $(r,\ta)\to(r_0,\ta_0)$ of the second factors in~\eqref{w3a} is still path dependent unless we take
\begin{equation}\label{w4}
    B_{,\ta\ta\ta}|_{(r_0,\ta_0)}=0.
\end{equation}
With this additional constraint, the limits, as $(r,\ta)\to(r_0,\ta_0)$, of the terms in~\eqref{w3a} have well defined values
\begin{align}\label{w3b}
&\lim_{(r,\ta)\to(r_0,\ta_0)}\frac{B_{,\ta}}{r-B}=\frac{B'_{,\ta}|_{(r_0,\ta_0)}}{1-B'|_{(r_0,\ta_0)}},\nn\\
&\lim_{(r,\ta)\to(r_0,\ta_0)}\frac{B_{,\ta\ta}}{r-B}=\frac{B'_{,\ta\ta}|_{(r_0,\ta_0)}}{1-B'|_{(r_0,\ta_0)}},
\end{align}
provided $B'|_{(r_0,\ta_0)}\neq 1$. If $B'|_{(r_0,\ta_0)}= 1$, we have to impose the following extra constraints
\begin{equation}\label{w3c}
B'_{,\ta}|_{(r_0,\ta_0)}=0,\quad B'_{,\ta\ta}|_{(r_0,\ta_0)}=0,\quad B'_{,\ta\ta\ta}|_{(r_0,\ta_0)}=0.
\end{equation}

Thus, we have shown that the curvature scalar is regular everywhere off the throat and it has a well defined and finite value on it if (a) the constraints~\eqref{w3} and~\eqref{w4} are satisfied in case $B'|_{(r_0,\ta_0)}\neq 1$ or (b) the constraints~\eqref{w3}, \eqref{w4}, and~\eqref{w3c} are satisfied in case $B'|_{(r_0,\ta_0)}= 1$. An instance of functions $B$ that satisfy the constraints~\eqref{w3}, \eqref{w4}, and~\eqref{w3c} are the one-variable relations $B\equiv B(r)$. In this case, the curvature scalar converges everywhere off the throat and on it whether $B'|_{(r_0,\ta_0)}\neq 1$ or not.

The mathematical expression of the Kretschmann scalar $R_{\al\bt\mu\nu}R^{\al\bt\mu\nu}$ is very sizeable, so we will not give here, however, we find that the only terms that may diverge on the throat are the following:
{\small\begin{align}\label{w4a}
&-\frac{9 \sin ^4\theta  B_{,\theta }{}^4}{4 r^4 K^4 (r-B)^4}\nn\\
&+\frac{3 \sin ^2\theta   \{\sin ^2\theta  [B_{,\theta \theta
}- (\ln  K)_{,\theta }B_{,\theta }]-\cos  \theta  B_{,\theta }\}B_{,\theta }{}^2}{r^4 K^4 (r-B)^3}\nn\\
&+\frac{[2 \cos ^2 \theta +3 \sin ^2 \theta  K
(r K)' (1-B')] B_{,\theta }{}^2}{r^4 K^4(r-B)^2}\nn\\
&+\frac{ \{2 N^2 [2 (\ln  K)_{,\theta }{}^2+(\ln  N)_{,\theta
}{}^2]-r^2 K^2 \sin ^2\theta  \omega _{,\theta }{}^2\} \sin ^4\theta  B_{,\theta }{}^2}{2 r^4 K^4N^2(r-B)^2}\nn\\
&+\frac{ [\sin ^2\theta  B_{,\theta
\theta }-2 \cos  \theta  B_{,\theta }-2 \sin ^2\theta   (\ln  K)_{,\theta }B_{,\theta }] \sin ^2\theta  B,_{\theta \theta }}{r^4 K^4(r-B)^2}\nn\\
&+ \frac{2 \sin ^2\theta  (1-B')[1+r
(\ln  K)'] B_{,\theta \theta }}{r^4 K^2 (r-B)}\\
&+\big[\frac{2 \sin ^2\theta  (\ln  N)' (\ln
 N)_{,\theta }}{r^3 K^2}-[1+r (\ln  K)'] \frac{4 \cos  \theta }{r^4 K^2}\nn\\
& -\frac{\sin ^4\theta  \omega ' \omega ,_{\theta }}{r N^2}\big]
\frac{(1-B') B_{,\theta }}{(r-B)}\nn\\
& +\big[\frac{13 (\ln  K)'}{ r^4K^2}+\frac{2 (\ln  K)'{}^2+2 (\ln  N)'{}^2}{r^3K^2}+\frac{6
K''}{r^3K^3}-\frac{1}{ r^5K^2}\nn\\
&-\frac{\sin ^2\theta  \omega '{}^2}{N^2 r}\big] \frac{\sin ^2\theta  B_{,\theta }{}^2}{(r-B)},\nn
\end{align}}%
and that all the other terms of $R_{\al\bt\mu\nu}R^{\al\bt\mu\nu}$ are finite on the throat and elsewhere. All that we have said in the paragraph following~\eqref{w3c}, concerning the convergence of the curvature scalar, applies to the convergence of the Kretschmann scalar.

We have derived the properties of~\eqref{w1} which were not discussed elsewhere. We refer the reader to Refs.~\cite{teo,Lobo} for an extended discussion of its other properties. It was particularly shown that the physical\footnote{If $k^{\mu}$ denotes a null vector, the condition $T_{\mu\nu}k^{\mu}k^{\nu}\geq 0$ is called the physical NEC. In general relativity, this implies the geometrical NEC $R_{\mu\nu}k^{\mu}k^{\nu}\geq 0$. In nonrotating solutions, the physical NEC implies the effective NEC $\ep+p_r\geq 0$ and $\ep+p_t\geq 0$, where $\ep$ is the energy density and ($p_r,p_t$) are the radial and transverse pressures, respectively. If $u^{\mu}$ denotes a timelike vector, similar definitions exist for the physical $T_{\mu\nu}u^{\mu}u^{\nu}\geq 0$ and geometrical $G_{\mu\nu}u^{\mu}u^{\nu}\geq 0$ WEC.} NEC, that is $T_{\mu\nu}k^{\mu}k^{\nu}\geq 0$ where $k^{\mu}$ is a null vector, is not violated in some regions around the throat allowing an infalling observer to avoid the (necessary) exotic matter sustaining the throat.

\section{First example of a rotating wormhole\label{frst}}

The first rotating wormhole derived using the metric~\eqref{w1} is Teo's wormhole~\cite{teo} given by
\begin{equation}\label{f1}
N=K=1+\frac{(4a\cos\ta)^2}{r},\quad B=1,\quad \om=\frac{2a}{r^3},
\end{equation}
where $a$ is the rotation parameter. Very recently, Teo's wormhole has been used to study collisional processes in the geometry of a rotating wormhole~\cite{col1,col2}.

For at least $a=1/4$, it was concluded~\cite{teo} that null and timelike geodesics, passing through the neck of the wormhole, do not encounter exotic matter; that it, they do not observe violations of the NEC.

Since $K$ depends on ($r,\ta$) while $B$ is constant, by the requirement~\eqref{g4} Teo's metric~\eqref{f1} cannot be derived from~\eqref{g2} by the coordinate transformation~\eqref{g23}.

It was shown in Ref.~\cite{W3} that Teo's wormhole cannot be generated by a single perfect or anisotropic fluid. It is, however, possible to show that this solution is generated by two rotating anisotropic fluids the SETs of which are unphysical and not related to any known matter distributions. In the following we outline the steps of the proof without providing the full expressions of the two SETs which are very sizeable.

The unphysical thing with Teo's wormhole is the component $G_{r\ta}$ of the Einstein tensor which is not zero
\begin{equation*}
G_{r\ta}=-\frac{32 a^2 (3 r+16 a^2 \cos^2\theta ) \cos\theta\sin\ta}{r (r+16 a^2 \cos ^2\theta )^2}.
\end{equation*}
Because of this property, we use a normal orthonormal basis $\mathbf{b}=(e_t,\,e_r,\,e_{\ta},\,e_{\vp})$
\begin{align}
&e^{\mu}_t=\Big(\frac{g_{tt}-2\om^2g_{\vp\vp}}{Ng_{tt}},0,0,-\frac{\om}{N}\Big),\nn\\
&e^{\mu}_r=\Big(0,\frac{\sqrt{r-B}}{\sqrt{r}},0,0\Big),\quad e^{\mu}_{\ta}=\Big(0,0,\frac{1}{rK},0\Big),\nn\\
\label{f2}&e^{\mu}_{\vp}=\Big(-\frac{2r\om K\sin\ta}{g_{tt}},0,0,\frac{1}{rK\sin\ta}\Big),
\end{align}
(where $g_{tt}=N^2-r^2\om^2K^2\sin^2\ta$, $g_{\vp\vp}=-r^2K^2\sin^2\ta$, and $g_{t\vp}=-\om g_{\vp\vp}$) in terms of which we expand the first anisotropic SET $T^{\mu\nu}$
\begin{equation}\label{f3}
    T^{\mu\nu}=\ep e^{\mu}_te^{\nu}_t+p_re^{\mu}_re^{\nu}_r+p_{\ta}e^{\mu}_{\ta}e^{\nu}_{\ta}+p_{\vp}e^{\mu}_{\vp}e^{\nu}_{\phi},
\end{equation}
where ($\ep, p_r,p_{\ta},p_{\vp}$) are the energy density and the pressure components of the first SET, along with a second orthonormal, but \emph{unusual}, basis $\mathbf{\bar{b}}=(\bar{e}_t,\,\bar{e}_r,\,\bar{e}_{\ta},\,\bar{e}_{\vp})$
\begin{align}
&\bar{e}^{\mu}_t=\Big(\frac{1}{N},0,0,\frac{\om}{N}\Big),\quad \bar{e}^{\mu}_{\vp}=\Big(0,0,0,\frac{1}{rK\sin\ta}\Big),\nn\\
&\bar{e}^{\mu}_r=\Big(0,\frac{r-B}{r},-\frac{\sqrt{B}}{r^{3/2}K},0\Big),\nn\\
\label{f4}&\bar{e}^{\mu}_{\ta}=\Big(0,\frac{\sqrt{B(r-B)}}{r},\frac{\sqrt{r-B}}{r^{3/2}K},0\Big),
\end{align}
in terms of which we expand the second anisotropic SET $\bar{T}^{\mu\nu}$
\begin{equation}\label{f5}
    \bar{T}^{\mu\nu}=\bar{\ep} \bar{e}^{\mu}_t\bar{e}^{\nu}_t+\bar{p}_r\bar{e}^{\mu}_r\bar{e}^{\nu}_r
    +\bar{p}_{\ta}\bar{e}^{\mu}_{\ta}\bar{e}^{\nu}_{\ta}+\bar{p}_{\vp}\bar{e}^{\mu}_{\vp}\bar{e}^{\nu}_{\vp},
\end{equation}
where ($\bar{\ep}, \bar{p}_r,\bar{p}_{\ta},\bar{p}_{\vp}$) are the energy density and the pressure components of the second SET.

The basis $\mathbf{\bar{b}}$ has been constructed so that $\bar{T}_{r\ta}\neq 0$. The nonvanishing components of $T_{\mu\nu}$ and $\bar{T}_{\mu\nu}$ are
\begin{align}
&T_{tt}=N^2\ep -\om^2g_{\vp\vp}p_{\vp},\quad T_{rr}=\frac{rp_r}{r-B},\nn\\
&T_{\ta\ta}=r^2K^2p_{\ta},\quad T_{t\vp}=\frac{g_{t\vp}[N^2(2\ep+p_{\vp})-\om^2g_{\vp\vp}p_{\vp}]}{g_{tt}},\nn\\
\label{f6}&T_{\vp\vp}=\frac{4N^2g_{t\vp}^2\ep -g_{\vp\vp}(N^2-\om^2g_{\vp\vp})^2p_{\vp}}{g_{tt}^2},
\end{align}
\begin{align}
&\bar{T}_{tt}=N^2\bar{\ep} -\om^2g_{\vp\vp}\bar{p}_{\vp},\quad \bar{T}_{r\ta}=\sqrt{rB}K(\bar{p}_{\ta}-\bar{p}_{r}),\nn\\
&\bar{T}_{rr}=\bar{p}_r+\frac{B\bar{p}_{\ta}}{r-B},\quad \bar{T}_{\ta\ta}=rK^2[(r-B)\bar{p}_{\ta}+B\bar{p}_{r}],\nn\\
\label{f7}&\bar{T}_{t\vp}=-g_{t\vp}\bar{p}_{\vp},\quad \bar{T}_{\vp\vp}=-g_{\vp\vp}\bar{p}_{\vp}.
\end{align}

We divide the field equations into two groups
\begin{multline}\label{f8a}
\text{\textbf{G1}}:\;G_{tt}=8\pi (T_{tt}+\bar{T}_{tt}),\quad G_{t\vp}=8\pi (T_{t\vp}+\bar{T}_{t\vp}),\\
G_{\vp\vp}=8\pi (T_{\vp\vp}+\bar{T}_{\vp\vp}),\qquad\qquad\qquad\quad\quad
\end{multline}
for the four unknowns ($\ep,p_{\vp},\bar{\ep},\bar{p}_{\vp}$) and
\begin{multline}\label{f8b}
\text{\textbf{G2}}:\;G_{rr}=8\pi (T_{rr}+\bar{T}_{rr}),\quad G_{r\ta}=8\pi \bar{T}_{r\ta},\\ G_{\ta\ta}=8\pi (T_{\ta\ta}+\bar{T}_{\ta\ta}),\qquad\qquad\qquad\quad\quad
\end{multline}
for the four unknowns ($p_r,p_{\ta},\bar{p}_r,\bar{p}_{\ta}$). Since in each group the number of unknowns exceeds the number of equations we can fix some unknowns and solve for the others.

Choosing $\ep$ of the form
\begin{equation}\label{f9}
\ep =\frac{q^2}{8\pi (rK)^4},
\end{equation}
which corresponds to an electromagnetic energy density, does not lead to any consistent (nonrotating or rotating) solution. Rather, we choose ($\ep,p_r$) of the form
\begin{multline}\label{f10}
\ep =-\frac{c_1^2}{64\pi (rK)^3},\quad p_r =-\frac{c_2^2}{64\pi (rK)^3},\\
\text{with}\quad 3<c_1^2<7\quad \text{ and }\quad 8<c_2^2<12,\quad\quad
\end{multline}
which correspond to an exotic matter. On substituting these values into \textbf{G1} and \textbf{G2} we derive the expressions of the unique remaining unknowns. Since their expressions, and the expressions of their series expansions in $a^2$, are sizeable we will not provide them here. Their series are of the form
\begin{align}
&p_{\ta}=\frac{12-c_2^2}{64\pi r^3}+\mathcal{O}(a^2),\quad p_{\vp}=\frac{c_1^2-3}{64\pi r^3}+\mathcal{O}(a^2),\nn\\
&\bar{\ep}=\frac{c_1^2}{64\pi r^3}+\mathcal{O}(a^2),\quad \bar{p}_r=\frac{c_2^2-8}{64\pi r^3}+\mathcal{O}(a^2),\\
&\bar{p}_{\ta}=\frac{c_2^2-8}{64\pi r^3}+\mathcal{O}(a^2),\quad \bar{p}_{\vp}=\frac{7-c_1^2}{64\pi r^3}+\mathcal{O}(a^2).\nn
\end{align}
For instance, up to $\mathcal{O}(a^4)$ the expression of $p_{\vp}$ reads
\begin{multline*}
p_{\vp}=\frac{c_1^2-3}{64\pi r^3}+\frac{9-3 [3+4 (6+c_1^2) r^3-24 r^4] \cos ^2\theta }{16 \pi  r^7} \,a^2\\+\mathcal{O}(a^4).\qquad\qquad\qquad\qquad\qquad\qquad\qquad\qquad
\end{multline*}

Thus, we have shown that Teo's rotating wormhole could be interpreted as a solution generated by two anisotropic fluids one of which, $T^{\mu\nu}$, is exotic and the other one, $\bar{T}^{\mu\nu}$, corresponds to ordinary matter. The corresponding nonrotating wormhole is also generated by two anisotropic fluids given by
\begin{align*}
&\text{exotic:}\;T^{\mu\nu}_{\text{s}}={\rm diag}\Big(\frac{-c_1^2}{64\pi r^3},\frac{-c_2^2}{64\pi r^3},\frac{12-c_2^2}{64\pi r^3},\frac{c_1^2-3}{64\pi r^3}\Big),\\
&\text{ordinary:}\;\bar{T}^{\mu\nu}_{\text{s}}={\rm diag}\Big(\frac{c_1^2}{64\pi r^3},\frac{c_2^2-8}{64\pi r^3},\frac{c_2^2-8}{64\pi r^3},\frac{7-c_1^2}{64\pi r^3}\Big),
\end{align*}
with a vanishing total energy density\footnote{The bases $\mathbf{b}$ and $\mathbf{\bar{b}}$ as well as the basis $\mathbf{\tilde{b}}$ introduced in Sec.~\ref{scnd} all coincide if rotation is suppressed. Hence, one can add, say, energy densities to find the total nonrotating density.}, a total radial pressure of $-1/(8\pi r^3)$, and a total transverse pressure of $-1/(16\pi r^3)$.

Based on different choices than~\eqref{f10}, other reinterpretations of Teo's wormhole remain possible due to the nonlinearity of the field equations; that is, the same metric may be sourced by different SETs.

In the following section we will construct a new rotating wormhole solution that is generated by an ordinary electromagnetic fluid and an exotic one, which are both anisotropic. A simple expression for its electromagnetic energy density would be given by~\eqref{f9}.

\section{Rotating wormhole with an electromagnetic charge\label{scnd}}

We keep using the basis~\eqref{f2}, the expansion~\eqref{f3} of the first SET and its nonvanishing components~\eqref{f6}, and we introduce a new basis $\mathbf{\tilde{b}}=(\tilde{e}_t,\,\tilde{e}_r,\,\tilde{e}_{\ta},\,\tilde{e}_{\vp})$ defined by,
\begin{align}
&\tilde{e}^{\mu}_t=\Big(\frac{1}{N},0,0,\frac{\om}{N}\Big)\,,\;\tilde{e}^{\mu}_r=\Big(0,\frac{\sqrt{r-B}}{\sqrt{r}},0,0\Big)\nn\\
\label{e4}&\tilde{e}^{\mu}_{\ta}=\Big(0,0,\frac{1}{rK},0\Big)\,,\;\tilde{e}^{\mu}_{\vp}=\Big(0,0,0,\frac{1}{rK\sin\ta}\Big),
\end{align}
and define a new SET $\tilde{T}^{\mu\nu}$
\begin{equation}\label{e5}
    \tilde{T}^{\mu\nu}=\tilde{\ep} \tilde{e}^{\mu}_t\tilde{e}^{\nu}_t+\tilde{p}_r\tilde{e}^{\mu}_r\tilde{e}^{\nu}_r
    +\tilde{p}_{\ta}\tilde{e}^{\mu}_{\ta}\tilde{e}^{\nu}_{\ta}+\tilde{p}_{\vp}\tilde{e}^{\mu}_{\vp}\tilde{e}^{\nu}_{\vp},
\end{equation}
with
\begin{align}
&\tilde{T}_{tt}=N^2\tilde{\ep} -\om^2g_{\vp\vp}\tilde{p}_{\vp},\quad \tilde{T}_{rr}=\frac{r\tilde{p}_r}{r-B},\nn\\
\label{e6}&\tilde{T}_{\ta\ta}=r^2K^2\tilde{p}_{\ta},\quad \tilde{T}_{t\vp}=-g_{t\vp}\tilde{p}_{\vp},\quad \tilde{T}_{\vp\vp}=-g_{\vp\vp}\tilde{p}_{\vp}.
\end{align}

The aim of this section is to derive an exact rotating wormhole solution sourced by an ordinary electromagnetic SET and sustained by an exotic matter. For that end, we first seek to impose the physical constraint $G_{r\ta}\equiv 0$. The general expression of $G_{r\ta}$ corresponding to~\eqref{w1} reads
\begin{align}
&-\frac{2 r (r-B) K^2 N^2}{\sin\ta}\,G_{r\ta}=2r (r-B) N^2 K_{,\theta } K' \nn\\
\label{e1}&+r^3 (r-B)K^4 \omega _{,\theta }\omega ' \sin ^2\theta\\
&+r K N \{2 (r-B) N_{,\theta } K'+N [B_{,\theta} K'-2 (r-B) K'_{,\theta }]\}\nn\\
&+K^2 N \{N B_{,\theta }+2 (r-B) N_{,\theta }+r [B_{,\theta } N'-2 (r-B) N'_{,\theta }]\}.\nn
\end{align}

It is not easy to handle analytically the nonlinear differential equation $G_{r\ta}\equiv 0$, so from now on we restrict ourselves to simple solutions where ($N,B,\om$) depend only on $r$
\begin{equation}\label{e2}
N\equiv N(r),\quad B\equiv B(r),\quad\om\equiv aW(r).
\end{equation}
This will help us to construct \emph{exact analytic} solutions in \emph{closed forms} with pretty expressions for the components of the SET. Exact solutions in closed forms are very useful for astrophysical applications~\cite{types} and computer simulations~\cite{Sgr}. As we shall see in section~\ref{gen}, where we deal with the general case, that the simplification ansatz~\eqref{e2} won't help anymore getting exact solutions in closed forms [to the differential equation~\eqref{ge2}]. Another advantage in employing the ansatz~\eqref{e2} is to ensure regularity everywhere of the curvature and Kretschmann scalars, as stated in the paragraph following~\eqref{w3c}, since in this case $B$ is independent of $\ta$. The ansatz~\eqref{e2} too ensures regularity on the axis of rotation.

It is easy to see that the constraints~\eqref{e2} reduce $G_{r\ta}\equiv 0$ to $K_{,\theta }K'-KK'_{,\theta }=0$ or, equivalently, to $K(r,\ta)\equiv F(r)H(\ta)$. But asymptotic flatness~\eqref{w1b} requires $H(\ta)\equiv 1$ leaving $K$ as a function of $r$ only, which we may take of the form
\begin{equation}\label{e3}
    K=1+a^2f(r)\qquad (f\to 0\text{ as }r\to\infty).
\end{equation}
This is conform to the symmetry requirement~\eqref{g4}.

We start with the case where $N$ is constant. Asymptotic flatness~\eqref{w1b} requires
\begin{equation}\label{e7}
    N\equiv 1.
\end{equation}
We divide the fields equations into two groups
\begin{multline}\label{e8}
\text{\textbf{G3}}:\;G_{tt}=8\pi (T_{tt}+\tilde{T}_{tt}),\quad G_{t\vp}=8\pi (T_{t\vp}+\tilde{T}_{t\vp}),\\
G_{\vp\vp}=8\pi (T_{\vp\vp}+\tilde{T}_{\vp\vp}),\qquad\qquad\qquad\quad\quad
\end{multline}
for the four unknowns ($\ep,p_{\vp},\tilde{\ep},\tilde{p}_{\vp}$) and
\begin{equation}\label{e9}
\text{\textbf{G4}}:\;G_{rr}=8\pi (T_{rr}+\tilde{T}_{rr}),\quad G_{\ta\ta}=8\pi (T_{\ta\ta}+\tilde{T}_{\ta\ta}),
\end{equation}
for the unknowns ($p_r+\tilde{p}_r,p_{\ta}+\tilde{p}_{\ta}$).

Since in each group the number of unknowns exceeds the number of equations we can fix some unknowns and solve for the others. We make the choice~\eqref{f9} for $\ep$
\begin{equation}\label{f9a}
\ep =\frac{q^2}{8\pi (rK)^4},
\end{equation}
where $q$ is an electric or a magnetic charge and $rK$ is the radial proper distance. For a massless solution, a corresponding simple expression for $B(r)$ is
\begin{equation}\label{f9b}
B=\frac{q^2}{r}.
\end{equation}

We first solve \textbf{G3}~\eqref{e8} for ($\tilde{\ep},\tilde{p}_{\vp},p_{\vp}$). The expansions of ($\tilde{\ep},\tilde{p}_{\vp},p_{\vp}$) in powers of $a^2$ are the following
\begin{align}
\label{e10}&\tilde{\ep}=-\frac{q^2}{4\pi r^4}+\mathcal{O}(a^2),\\
&\tilde{p}_{\vp}=\frac{r [r (q^2-r^2) W''+(3 q^2-4 r^2) W']+8 q^2 W}{32 \pi  r^4 W}+\mathcal{O}(a^2),\nn\\
&p_{\vp}=-\frac{r [r (q^2-r^2) W''+(3 q^2-4 r^2) W']+4 q^2 W}{32 \pi  r^4 W}+\mathcal{O}(a^2),\nn
\end{align}
where the terms independent of $a$ (the leading terms) are the static values. We require that the leading terms of the SETs be independent of the choice of $W(r)$ by setting
\begin{equation}\label{e11}
    r [r (q^2-r^2) W''+(3 q^2-4 r^2) W']+8 q^2 W=0.
\end{equation}
This requirement sets to $0$ the leading term of $\tilde{p}_{\vp}$, which may correspond to a dust. Using~\eqref{e11} in~\eqref{e10}, we see that the leading term of $p_{\vp}$, $q^2/(4\pi r^4)$, is also independent of $W$ as required by our hypothesis that the SET $T^{\mu\nu}$ is that of an ordinary electromagnetic source. We will establish below that the leading terms of ($p_r,p_{\ta}$) are also independent of $W$ and correspond to an ordinary electromagnetic source.

The function $\om=aW$ is subject to the requirement that asymptotically it approach the angular velocity of a star, usually taken as $2a/r^3$ [compare with~\eqref{f1}]
\begin{equation}\label{e11a}
\om\to 2a/r^3\;\text{ as }\;r\to\infty.
\end{equation}
The differential equation~\eqref{e11} along with the boundary condition $W\to 2/r^3$ as $r\to\infty$ lead to the unique solution
\begin{widetext}
\begin{equation}\label{e12}
W=\frac{3 \e^{-\sqrt{7} \text{arccot}\,(\sqrt{X^2-1})} \big[\sqrt{X^2-1}+\sqrt{7}-(\sqrt{X^2-1}-\sqrt{7})\e^{2 \sqrt{7} \text{arccot}\,(\sqrt{X^2-1})}\big]}{8\sqrt{7} |q^3| X}\qquad (X\equiv r/|q|\geq 1),
\end{equation}
\end{widetext}
where $y=\text{arccot}\,x$ is the inverse function of $y=\cot x$ and $0<x<\pi$. In the expression of $W$ one may replace $\text{arccot}\,(\sqrt{X^2-1})$ by $\arctan (1/\sqrt{X^2-1})$ where $y=\arctan x$ is the inverse function of $y=\tan x$ and $-\pi/2<x<\pi/2$.

The sign of $W$ is that of the expression inside the square brackets in~\eqref{e12}, which we rewrite in terms of $Y\equiv \sqrt{X^2-1}$
\begin{multline}\label{e13}
S(Y)=Y+\sqrt{7}-(Y-\sqrt{7})\e^{2 \sqrt{7} \text{arccot}\,Y},
\\\text{where } Y\equiv \sqrt{X^2-1}\geq 0.\qquad\qquad
\end{multline}
Since $S(0)>0$ ($\text{arccot}\,0=\pi/2$), $\lim_{Y\to\infty}S\to 0^+$ [$S=16\sqrt{7}/(3Y^2)+\mathcal{O}(1/Y^3)$], and $S_{,YY}=32\sqrt{7}\e^{2 \sqrt{7} \text{arccot}\,Y}/(Y^2+1)^2>0$, we conclude that $S(Y)$ is always positive, and so is the function $W$. The latter may be written in the form
\begin{multline}
W=\frac{3 [\sqrt{7} \cosh (\sqrt{7} \text{arccot}\,Y)-Y \sinh (\sqrt{7} \text{arccot}\,Y)]}{4\sqrt{7}|q^3|\sqrt{Y^2+1}},
\\\text{with } Y\geq 0.\qquad\qquad\qquad\qquad\qquad\qquad
\end{multline}

With $f(r)\neq 0$~\eqref{e3}, the expressions of the SETs are still sizeable. If $f(r)= 0$, the rotating wormhole and its corresponding nonrotating one read, respectively
\begin{equation}\label{e14}
    \dd s^2=\dd t^2-\frac{\dd r^2}{1-\frac{q^2}{r^2}}-r^2[\dd\ta^2+\sin^2\ta(\dd\varphi-aW\dd t)^2],
\end{equation}
\begin{equation}\label{e14}
    \dd s_{\text{s}}^2=\dd t^2-\frac{\dd r^2}{1-\frac{q^2}{r^2}}-r^2(\dd\ta^2+\sin^2\ta\dd\varphi^2).
\end{equation}
The resolution of  the \textbf{G4}~\eqref{e9} provides the expressions of $p_r+\tilde{p}_r$ and $p_{\ta}+\tilde{p}_{\ta}$; then, it is matter of comparison (with the nonrotating case) and identification to extract the expressions of ($p_r,p_{\ta}$) which correspond to an ordinary electromagnetic SET\footnote{It is also a matter of choice: we could modify both ($p_r,\tilde{p}_r$) without modifying their sum and we could do the same for ($p_{\ta},\tilde{p}_{\ta}$). Only the total values of the pressures $p_r+\tilde{p}_r$ and $p_{\ta}+\tilde{p}_{\ta}$ are determined analytically and the partial pressures are determined, say, by the wormhole ``assembler" (from an advanced civilization).}. Finally, the energy densities and the pressures of the two fluids for the rotating wormhole are given by
\begin{align}
\label{e15a}&\ep =-p_r=p_{\ta}=\frac{q^2}{8\pi r^4},\\
&p_{\vp}=\frac{q^2 (1-5 a^2 r^2  W^2\sin ^2\theta+2 a^4 r^4  W^4\sin ^4\theta )}{8 \pi  r^4 (1+a^2 r^2 W^2\sin ^2\theta)},\nn
\end{align}
\begin{align}
&\tilde{\ep}=-\frac{q^2}{4 \pi  r^4}-\frac{a^2 q^2 r^2  W^2\sin ^2\theta }{\pi  r^4 (1+a^2 r^2 W^2\sin ^2\theta )}\nn\\
\label{e15b}&\quad+\frac{a^2 (q^2-r^2) W'^{\,2}\sin ^2\theta }{32 \pi},\\
&\tilde{p}_{\ta}=-\tilde{p}_r=\frac{a^2 (q^2-r^2) W'^{\,2}\sin ^2\theta }{32 \pi },\nn\\
&\tilde{p}_{\vp}=-\frac{a^2 [8 q^2 W^2-3 r^2 (q^2-r^2) W'^{\,2}] \sin ^2\theta }{32 \pi  r^2},\nn
\end{align}
where we have used the differential equation~\eqref{e11} to eliminate $W''$ from the expressions of ($p_{\vp},\tilde{\ep},\tilde{p}_{\vp}$). The static values are obtained setting $a=0$
\begin{align}
&\ep_{\text{s}} =-p_{r\,\text{s}}=p_{\ta\,\text{s}}=p_{\vp\,\text{s}}=\frac{q^2}{8\pi r^4},\nn\\
\label{e16}&\tilde{\ep}_{\text{s}}=-\frac{q^2}{4 \pi  r^4},\quad \tilde{p}_{r\,\text{s}}=\tilde{p}_{\ta\,\text{s}}=\tilde{p}_{\vp\,\text{s}}=0.
\end{align}
This shows that the nonrotating and rotating wormholes are generated by the SET of an ordinary electromagnetic field and by that of an exotic dust. Due to rotation, the two SETs become anisotropic.

The decomposition of the total SET of a nonrotating wormhole into an ordinary electromagnetic (a source-free radial electric or magnetic) part and an exotic one~\eqref{e16} was considered in Ref.~\cite{snk,ex,az3,BK}. The static solution is just the so-called Ellis wormhole~\cite{BE1,Ellis} as this can be seen performing the radial coordinate change $r^2=u^2+q^2$, which is the same as $u=|q|Y$, yielding
\begin{align}
\label{ss}&\dd s^2=\dd t^2-\dd u^2-(u^2+q^2)[\dd\ta^2+\sin^2\ta(\dd\varphi-aW\dd t)^2],\\
&\dd s_{\text{s}}^2=\dd t^2-\dd u^2-(u^2+q^2)(\dd\ta^2+\sin^2\ta\dd\varphi^2).
\end{align}

The axially symmetric solution given by Eqs.~(41), (46), and~(47) of Ref.~\cite{az3}, which is sourced by two rotating fluids one of which is electromagnetic and the other one is exotic, has been interpreted as a rotating wormhole with no dragging effects ($\om\equiv 0$). One could interpret it as a nonrotating wormhole sourced by two rotating fluids, in which case this would generalize the Bronnikov-Ellis wormhole. The sought rotating wormhole with dragging effects, a counterpart of the Bronnikov-Ellis wormhole, is the one given in~\eqref{ss}. This seems to be the simplest rotating Bronnikov-Ellis wormhole; other rotating Bronnikov-Ellis wormholes are possible in general relativity as well as in generalized theories of gravity~\cite{KK}.

\section{The general case\label{gen}}

In this section we treat the general case of a rotating wormhole where ($N,B,\om$) are any functions of $r$, as in~\eqref{e2}, with the energy density of one of the two fluids, $T^{\mu\nu}$, being electromagnetic given by~\eqref{f9a}. Without loss of generality, we take $K\equiv 1$ as in the nonrotating solution. We then specialize to the case of a massive wormhole with an electromagnetic charge.

\subsection{The nonrotating wormhole}

The field equations governing the nonrotating wormhole with metric
\begin{equation}\label{s1}
    \dd s^2=N^2\dd t^2-\frac{\dd r^2}{1-\frac{B}{r}}-r^2(\dd\ta^2+\sin^2\ta\dd\varphi^2),
\end{equation}
read
\begin{align}\label{s2}
&B'=8\pi r^2(\ep_{\text{s}} +\tilde{\ep}_{\text{s}}) ,\nn\\
&2(\ln N)'=\frac{8 \pi  r^3(p_{r\,\text{s}}+\tilde{p}_{r\,\text{s}}) +B}{r (r-B)},\\
&2 (p_{t\,\text{s}}+\tilde{p}_{t\,\text{s}})=2 (p_{r\,\text{s}}+\tilde{p}_{r\,\text{s}})+r (p_{r\,\text{s}}+\tilde{p}_{r\,\text{s}})'\nn\\
&\quad\quad\quad\quad\quad\quad +r(p_{r\,\text{s}}+\tilde{p}_{r\,\text{s}}+\ep_{\text{s}} +\tilde{\ep}_{\text{s}} )(\ln N)',\nn\\
&\text{with }\ep_{\text{s}} =-p_{r\,\text{s}}=\frac{q^2}{8\pi r^4},\nn
\end{align}
where the subscript ``$t$" denotes a transverse pressure: $p_{t\,\text{s}}=p_{\ta\,\text{s}}=p_{\vp\,\text{s}}$ and $\tilde{p}_{t\,\text{s}}=\tilde{p}_{\ta\,\text{s}}=\tilde{p}_{\vp\,\text{s}}$.
($\ep_{\text{s}},p_{r\,\text{s}}$) have the same expressions as in~\eqref{e16}.

In the case of a massive wormhole we have $\lim_{r\to\infty}B=2M$ with $M$ being the mass. As we saw in the previous section (Sec.~\ref{scnd}), the case of $N$ being constant ($N\equiv 1$) could be supported by a dust SET $\tilde{T}^{\mu\nu}$ if the mass of the solution is null without setting constraints on the values of the other parameters. If $M\neq 0$, assuming $\tilde{T}^{\mu\nu}$ to be the SET of a dust, the case $N\equiv 1$ would lead to restriction(s) on the parameters' values. For instance, taking $\tilde{p}_{r\,\text{s}}=0$, the second line~\eqref{s2} evaluated at the throat $r_0$ [$r_0=B(r_0)$] implies $8 \pi  r_0^3p_{r\,\text{s}}(r_0) +B(r_0)=0$ or, equivalently, $q^2=r_0^2$. In the remaining part of this section, we will neither restrict ourselves to the case $N\equiv 1$ nor to the case where $\tilde{T}^{\mu\nu}$ is the SET of a dust. We will however impose the asymptotic behavior~\eqref{w1b}
\begin{equation}\label{ab1}
    N\sim 1-\frac{N_{\infty}}{r^{\al}}\;\text{ as }\;r\to\infty\qquad (\al>0).
\end{equation}

For any physical wormhole solution, the SET vanishes at spatial infinity. For a nonrotating solution we may write asymptotically
\begin{equation}\label{ab2}
    P_{r\,\text{s}}\sim \frac{P_{\infty}}{r^{\bt}}\;\text{ as }\;r\to\infty\qquad (\bt>0),
\end{equation}
where $P_{r\,\text{s}}=p_{r\,\text{s}}+\tilde{p}_{r\,\text{s}}$ is the total radial pressure. In order to observe the flatness condition~\eqref{ab1} the second line~\eqref{s2} implies $\bt>2$. If, further, $\bt>3$ then $\al=1$ and $N_{\infty}=M$.

Based on their asymptotic behavior, nonrotating wormholes have been classified into three types~\cite{types}. If the total energy density $E_{\text{s}}=\ep_{\text{s}} +\tilde{\ep}_{\text{s}}$ behaves as
\begin{equation}\label{ab3}
    E_{\text{s}}\sim \frac{E_{\infty}}{r^{\gamma}}\;\text{ as }\;r\to\infty\qquad (\gamma>0),
\end{equation}
the classification yields~\cite{types}
\begin{description}
  \item[] type I: $\bt - \ga > 1$$\qquad (\bt >3,\ga >3)$;
  \item[] type II: $0<\bt - \ga \leq 1$$\qquad (\bt >3,\ga >3)$;
  \item[] type III: $\bt \leq \ga$$\qquad (\bt >3,\ga >3)$.
\end{description}
Type I (respectively type III) wormholes violate the least (respectively the most) the LECs.

Among the nonrotating type I wormholes derived in Ref.~\cite{types}, we select the solution having a positive total energy density $E_{\text{s}}=\ep_0r_0^4/r^4$ given  by
\begin{align}\label{s3a}
&B=2M-\frac{(2M-r_0)r_0}{r}\qquad \Big(\frac{r_0}{2}<M<r_0\Big),\nn\\
&N^2=\exp\Big(-\sum_{i=1}^{n-3}\frac{S_i}{i\,y^{i}}\Big)<1\qquad [n \text{ (integer)}\geq 6],
\end{align}
\begin{align}\label{s3b}
&\ep_{\text{s}} =\frac{q^2}{8\pi r^4},\quad  \tilde{\ep}_{\text{s}}=\frac{\ep_0r_0^4}{r^4}-\frac{q^2}{8\pi r^4},\nn\\
&p_{r\,\text{s}}=-p_{t\,\text{s}}=-\frac{q^2}{8\pi r^4}, \\
&\tilde{p}_{r\,\text{s}}=-\frac{(y-1)S_{n-2}+1}{8\pi r_0^{2}y^{n+1}}+\frac{q^2}{8\pi r^4},\nn\\
&\tilde{p}_{t\,\text{s}}=\frac{2[(n-2)y+1-n]S_{n-2}+2(n-1)}{32\pi r_0^2y^{n+1}}\nn\\
&\quad +\frac{[xy^{n-3}-(y-1)S_{n-2}-1](\sum_{i=1}^{n-3}\frac{S_i}{y^{i}})}{32\pi r_0^2y^{n+1}}-\frac{q^2}{8\pi r^4},\nn
\end{align}
where $\ep_0>0$ is the total energy density $E_{\text{s}}(r_0)$ at the throat $r_0$ and $M$ is the mass of the wormhole
\begin{equation}\label{s4}
2M=r_0+8\pi r_0^3\ep_0.
\end{equation}
The constraint $r_0/2<M$ results from the positiveness of total energy density~\cite{types} and the constraint $M<r_0$ results from $B'(r_0)<1$~\cite{types}. In~\eqref{s3a} and~\eqref{s3b}, ($x,y,S_k$) are defined as follows:
\begin{align}\label{s4}
&x\equiv (2M-r_0)/r_0\qquad (0<x<1),\nn\\
&y\equiv r/r_0,\\
&S_k\equiv \sum_{i=0}^kx^i=\frac{1-x^{k+1}}{1-x}.\nn
\end{align}

Since $0\neq\tilde{p}_{r\,\text{s}}\neq \tilde{p}_{t\,\text{s}}\neq 0$~\eqref{s3b}, the exotic SET $\tilde{T}^{\mu\nu}$ does not correspond to a dust; rather, it corresponds to an anisotropic fluid.

Notice that in~\eqref{s2} and~\eqref{s3b}, the expressions of ($p_{r\,\text{s}}+\tilde{p}_{r\,\text{s}},p_{t\,\text{s}}+\tilde{p}_{t\,\text{s}}$) could be arranged as follows:
\begin{align}
\label{s5}&p_{r\,\text{s}}+\tilde{p}_{r\,\text{s}}=\frac{2 r (r-B) N'-B N}{8 \pi  r^3 N},\\
\label{s6}&p_{t\,\text{s}}+\tilde{p}_{t\,\text{s}}=\frac{B [N-r (N'+2 r N'')]}{16 \pi
r^3 N}\nn\\
&\quad\quad\quad\quad -\frac{B' N -r [(2-B') N'+2 r N'']}{16 \pi r^2 N}.
\end{align}

\subsection{The rotating wormhole}

While the expressions of ($N,B$) have been fixed in~\eqref{s3a}, the following treatment is valid for any functions ($N,B,\om$) of $r$ as in~\eqref{e2}.

This time we first solve \textbf{G4}~\eqref{e9} for ($p_r+\tilde{p}_r,p_{\ta}+\tilde{p}_{\ta}$). Their expansions in powers of $a^2$ are
\begin{align}
\label{ge3}&p_r+\tilde{p}_r=\text{rhs of~\eqref{s5}}+\mathcal{O}(a^2),\\
&p_{\ta}+\tilde{p}_{\ta}=\text{rhs of~\eqref{s6}}+\mathcal{O}(a^2).\nn
\end{align}
Notice that the leading terms of ($p_r+\tilde{p}_r,p_{\ta}+\tilde{p}_{\ta}$) do not depend on any \emph{choice} of $W$.

Now, we solve \textbf{G3}~\eqref{e8} for ($\tilde{\ep},\tilde{p}_{\vp},p_{\vp}$). The expansion of $\tilde{\ep}$ in powers of $a^2$ is
\begin{equation}\label{ge0}
\tilde{\ep}=-\frac{q^2-r^2B'}{8\pi r^4}+\mathcal{O}(a^2),
\end{equation}
the leading term of which is also independent of any \emph{choice} of $W$: this is precisely $\tilde{\ep}_{\text{s}}$~\eqref{s3b}. The leading terms of the expansions of ($p_{\vp},\tilde{p}_{\vp}$) in powers of $a^2$ do, however, depend on the choice of $W$. If the latter is chosen to satisfy the differential equation:
\begin{multline}\label{ge2}
2 r^3 (r-B) N W''-r^2 \{[7 B-r (8-B')] N+2 r (r-B) N'\} W'\\
 -16 q^2 N W=0,\qquad\qquad\qquad\qquad\qquad\qquad
\end{multline}
then the leading terms of the expansions of ($p_{\vp},\tilde{p}_{\vp}$) in powers of $a^2$ no longer depend on $W$ and take the forms
\begin{align}
\label{ge1}&p_{\vp}=\frac{q^2}{8\pi r^4} +\mathcal{O}(a^2),\\
&\tilde{p}_{\vp}=\text{rhs of~\eqref{s6}}-\frac{q^2}{8\pi r^4}+\mathcal{O}(a^2),\nn
\end{align}
which are precisely $p_{t\,\text{s}}$ and $\tilde{p}_{t\,\text{s}}$~\eqref{s3b}, respectively.

We have checked that the case $K\neq 1$ is no loss of generality; for instance, the leading term of $\tilde{\ep}$ remains also independent of any \emph{choice} of $W$, as in~\eqref{ge0}, but it depends on ($K,K',K''$) as the differential equation governing the behavior of $W$ does in this case too.

Finally, the energy densities and the pressures of the two fluids for the rotating wormhole are given by
\begin{align}
\label{ge3a}&\ep =-p_r=p_{\ta}=\frac{q^2}{8\pi r^4},\\
&p_{\vp}=\frac{q^2 (N^4-5 a^2 r^2 N^2 W^2 \sin ^2\theta +2 a^4 r^4 W^4 \sin ^4\theta )}{8 \pi  r^4 N^2 (N^2+a^2 r^2 W^2 \sin^2\theta )},\nn
\end{align}
\begin{align}
&\tilde{\ep}=\frac{B'}{8 \pi  r^2}-\frac{q^2 (N^2+9 a^2 r^2 W^2 \sin ^2\theta )}{8 \pi  r^4 (N^2+a^2 r^2 W^2 \sin ^2\theta)}\nn\\
\label{ge3b}&\qquad -\frac{a^2 r (r-B) W'^2 \sin ^2\theta }{32 \pi  N^2},\\
&\tilde{p}_r=\text{rhs of~\eqref{s5}}+\frac{q^2}{8\pi r^4}+\frac{a^2 r (r-B) W'^2 \sin ^2\theta }{32 \pi  N^2},\nn\\
&\tilde{p}_{\ta}=\text{rhs of~\eqref{s6}}-\frac{q^2}{8\pi r^4}-\frac{a^2 r (r-B) W'^2 \sin ^2\theta }{32 \pi  N^2},\nn\\
&\tilde{p}_{\vp}=\text{rhs of~\eqref{s6}}-\frac{q^2}{8\pi r^4}-\frac{a^2 q^2 W^2 \sin ^2\theta }{4 \pi  r^2 N^2}\nn\\
&\qquad -\frac{3 a^2 r (r-B) W'^2 \sin ^2\theta }{32 \pi  N^2},\nn
\end{align}
where we have used the differential equation~\eqref{ge2} to eliminate $W''$ from the expressions of ($p_{\vp},\tilde{\ep},\tilde{p}_{\vp}$). Since only the total values of the pressures $p_r+\tilde{p}_r$ and $p_{\ta}+\tilde{p}_{\ta}$ are determined analytically, here again, as was the case treated in~\eqref{e15a}, we fix ($p_r,p_{\ta}$) to their static values.

There is no way to solve~\eqref{ge2} in the general case where ($N,B$) are any functions as in the case where ($N,B$) are given by~\eqref{s3a}. Searching for a power series solution in $1/r$ satisfying the boundary condition~\eqref{e11a}, the expansion takes the form
\begin{equation}\label{ge4}
W=\frac{2}{r^3}+\frac{c_4}{r^4}+\frac{c_5}{r^5}+\frac{c_6}{r^6}+\mathcal{O}(1/r^7).
\end{equation}
The case to which one is mostly interested is the one with $\bt>3$~\eqref{ab2}, which results in $N_{\infty}=M$ as we saw earlier~\eqref{ab1}. This constraint on $\bt$ yields
\begin{equation}\label{ge4}
    c_4=0.
\end{equation}
Because of this, the dragging effects of these rotating wormholes mimic to some extent those of rotating stars\footnote{The solution given in~\eqref{e12} is also endowed with such a property. Its series expansion has only odd terms:
\begin{equation*}
W=\frac{2}{r^3}+\frac{11q^2}{5r^5}+\frac{253q^4}{140r^7}+\mathcal{O}(1/r^9).
\end{equation*}
}. The distinction is rendered possible only in the regions surrounding the throat.

The values of the other constants, $c_i$ and $i\geq 5$, depend on the particular nonrotating wormhole ($B,N$). If the latter is given by~\eqref{s3a}, taking $n=6$ the function $W$ for the corresponding rotating wormhole is approximated by the series
\begin{equation}\label{ge5}
W=\frac{2}{r^3}+\frac{16q^2-3(2M-r_0)r_0}{10r^5}\,\Big(1+\frac{10M}{9r}\Big)+\mathcal{O}(1/r^7).
\end{equation}
If we impose the constraint on the still-free parameters ($M,q,r_0$)
\begin{equation}\label{ge6}
16q^2=3(2M-r_0)r_0,
\end{equation}
this yields $c_5=c_6=0$. An observer falling into the geometry of these rotating wormholes will almost not be able to distinguish their dragging effects from those of rotating, non-flattened, stars except very near the throat. The solution reads
\begin{multline}\label{ge7}
W=\frac{2}{r^3}+3\frac{(2M-r_0)^4+4Mr_0(M-r_0)^2+4M^3r_0}{28r^7}\\+\mathcal{O}(1/r^8),\qquad\qquad\qquad\qquad\qquad\qquad
\end{multline}
where $c_7$, the coefficient of $1/r^7$, is manifestly positive. The constraint~\eqref{ge6} may be realized by the wormhole ``assembler" in different ways on observing the inequalities on $M$~\eqref{s3a}.

\section{The energy conditions\label{LEC}}

\subsection{The null energy condition}

The NEC is the constraint
\begin{equation}\label{n1}
    I_{\text{NEC}}\equiv (T_{\mu\nu}+\tilde{T}_{\mu\nu})k^{\mu}k^{\nu}\geq 0\qquad (\text{physical NEC})
\end{equation}
expressing the positiveness of the local energy as seen by any null vector $k^{\mu}$. Using the basis $\mathbf{\tilde{b}}$~\eqref{e4}, this is given by
\begin{equation}\label{n2}
    k^{\mu}=\tilde{e}^{\mu}_t+s_1\tilde{e}^{\mu}_r+s_2\tilde{e}^{\mu}_{\ta}+s_3\tilde{e}^{\mu}_{\vp}\qquad (s_1^2+s_2^2+s_3^2=1).
\end{equation}
Using the SET expressions~\eqref{f6} and~\eqref{e6}, and the basis $\mathbf{\tilde{b}}$~\eqref{e4} along with the general rotating metric~\eqref{w1} (taking $K\equiv 1$), we find
\begin{align}\label{n3}
&I_{\text{NEC}}=\ep+\tilde{\ep}+s_1^2(p_r+\tilde{p}_r)+s_2^2(p_{\ta}+\tilde{p}_{\ta})+s_3^2(p_{\vp}+\tilde{p}_{\vp})\nn\\
&\quad\quad\quad -\frac{4arNW\sin\ta}{g_{tt}^2}\Big[arNW(1+s_3)^2\sin\ta\\
&\quad\quad\quad +(N-arW\sin\ta)^2s_3\Big](\ep+p_{\vp}),\nn
\end{align}
which reduces to the nonrotating value if $a=0$.

It is easy to show that the NEC is violated by making special choices of ($s_1,s_2,s_3$). We will reach that conclusion below but our main purpose is to show that there are some paths whose tangent is $k^{\mu}$ along which no negative energy densities are noticed. Let us apply~\eqref{n3} to conical light paths, that is, to paths moving on the cone of equation $\ta =\text{constant}$ ($s_2\equiv 0$) from spatial infinity to the throat ($s_1<0$) and revolving in the same or opposite direction as the rotating wormhole ($s_3>0$ or $s_3<0$, respectively, with $|s_3|<1$). For such paths, $k^{\mu}$ takes the following form where $s_3$ is kept arbitrary and $s_1=-\sqrt{1-s_3^2}$
\begin{equation}\label{n4}
k^{\mu}=\Big(\frac{1}{N},-\frac{\sqrt{1-s_3^2}\sqrt{r-B}}{\sqrt{r}},0,\frac{aW}{N}+\frac{s_3}{r\sin\ta}\Big).
\end{equation}

In our first application we consider the rotating wormhole given by $N=1$, \eqref{e12}, \eqref{e15a} and\eqref{e15b}. We find
\begin{multline}\label{n5}
8\pi r^4I_{\text{NEC}}=[2q^2-a^2r^4(r^2-q^2)W'^2\sin^2\ta]s_3^2\\
+(8aq^2rW\sin\ta)s_3-2q^2.\quad\quad\quad\quad
\end{multline}
Note that at the throat $r_0=|q|$ we have $\lim_{r\to |q|}(r^2-q^2)W'^2=[36/(7q^6)]\sinh^2(\sqrt{7}\pi/2)\neq 0$. The rhs of~\eqref{n5} may have both signs for arbitrary $a$. However, the rotation parameter $a$ is subject to the constraint that the linear velocity at the throat, in the plane $\ta=\pi/2$, is much smaller than unity
\begin{equation}\label{n6}
    ar_0W(r_0)\ll 1.
\end{equation}
This is the slow rotation limit~\cite{ks,gh} ensuring that linear velocities of dragged objects do not exceed the speed of light to ensure safe traversability. This implies $arW(r)\ll 1$ since $rW(r)$ is a decreasing function of $r$. In this limit, the rhs of~\eqref{n5} has two roots $s_{3-}<-1$ and $0<s_{3+}<1$, which is the desired root given by
\begin{equation}\label{n7}
s_{3+}=1-2arW(r)\sin\ta +\mathcal{O}(a^2).
\end{equation}
Thus, the physical NEC, $(T_{\mu\nu}+\tilde{T}_{\mu\nu})k^{\mu}k^{\nu}\geq 0$, is not violated along such conical light paths satisfying $1>s_3\geq s_{3+}$ and $s_1=-\sqrt{1-s_3^2}$. In the case of no rotation, the rhs of~\eqref{n5} becomes $-2q^2s_1^2$ which is always negative unless we take $s_1\equiv 0$ corresponding to circular paths on the cone never reaching the throat. If no rotation, the physical NEC is not violated on conical circular paths and violated on any path that might cross the throat.

There is no ergosphere in the slow rotation limit~\eqref{n6} since in this case $g_{tt}=1-a^2r^2W^2\sin^2\ta>0$.

Now, consider the rotating wormhole given by~\eqref{s3a}, \eqref{s4}, \eqref{ge3a}, and~\eqref{ge3b}. Equation~\eqref{n3} reduces to
\begin{align*}
&16\pi r^3N^2I_{\text{NEC}}=\{N [N (3 B-r B')+2 r^2 N'' (r-B)\\
&+N' (3 r B -2 r^2-r^2 B')]-2 a^2 r^4 W'^2 (r-B) \sin ^2\theta \}s_3^2\\
&+16a q^2 N W \sin  \theta  s_3+2 N [N (r B'-B)+2 r N'(r-B)].
\end{align*}
Fixing $n=6$, this reads
\begin{equation}\label{n9}
16\pi r^3I_{\text{NEC}}=(A_1-2a^2A_2)s_3^2+aCs_3+D_1,
\end{equation}
where $D_1=16\pi r^3(E_{\text{s}}+P_{r\,\text{s}})$, $A_1=16\pi r^3(P_{t\,\text{s}}-P_{r\,\text{s}})$ ($P_{t\,\text{s}}\equiv p_{t\,\text{s}}+\tilde{p}_{t\,\text{s}}$ is the total transverse pressure of the nonrotating wormhole), $A_2$, and $C$ are given by
\begin{align}\label{n10}
&A_1=\frac{r_0x S_1^2 (1+x^2)^2}{2 y^7}-\frac{r_0S_1 (1+x^2) (1+x^4)}{2 y^6}-\frac{r_0S_6}{2 y^5}\nn\\
&+\frac{r_0x S_1}{2 y^2}-\frac{r_0S_1 [1+x (14+x)(1+x^2)]}{2 y^4}\nn\\
&+r_0\frac{12+x \{13+x [13+x (13+12 x)]\}}{2 y^3},\\
&A_2=\frac{r^4 W'^2 (r-B) \sin ^2\theta}{N^2}>0,\quad C=\frac{16 q^2 W \sin  \theta}{N}>0, \nn\\
&D_1=r_0\frac{2 S_1 (x+x^3)-2 S_4y+2 xy^3}{y^4},\nn
\end{align}
with $r-B=r_0(y-1)(y-x)/y$.

Numerically, we have checked that the term $A_1$ is always positive for all possible values of $x$~\eqref{s4}; $D_1$ is positive for all $x$ and $y>y_1>1$ and negative for all $x$ and $1\leq y<y_1$ where $y_1$ is the unique real root of $D_1=0$ as was established in~\cite{types}. Thus, in the slow rotation limit~\eqref{n6} the physical NEC is not violated along the conical light paths~\eqref{n4} if either $y\geq y_1$ ($s_3$ arbitrary) or $1\leq y<y_1$ and $-1<s_3\leq s_{3-}$ or $1>s_3\geq s_{3+}$ where ($s_{3-},s_{3+}$) are given by
\begin{equation}\label{n10}
s_{3\pm }=\pm \sqrt{\frac{-D_1}{A_1}}-\frac{aC}{2A_1}+\mathcal{O}(a^2)\qquad (1\leq y<y_1).
\end{equation}
The expression inside the square root is smaller than 1 for all possible values of $x$ if $1\leq y<y_1$. In case of no rotation, the conclusion remains valid with ($s_{3-},s_{3+}$) still given by~\eqref{n10} taking $a\equiv 0$.

For the nonrotating wormhole~\eqref{s3a}, \eqref{s4} (taking $n=6$) as well as for its rotating counterpart derived in this work, we conclude that if it were possible to direct light paths along the $r$-depend null vector~\eqref{n4}, no violation of the NEC at any event on these paths would be observed.

\subsection{The weak energy condition}

The WEC is the constraint
\begin{equation}\label{we1}
    I_{\text{WEC}}\equiv (T_{\mu\nu}+\tilde{T}_{\mu\nu})u^{\mu}u^{\nu}\geq 0\qquad (\text{physical WEC})
\end{equation}
expressing the positiveness of the local energy as seen by any timelike vector $u^{\mu}$ ($u^{\mu}u_{\mu}=1$). Using the basis $\mathbf{\tilde{b}}$~\eqref{e4}, this is given by
\begin{multline}\label{we2}
    u^{\mu}=U(\tilde{e}^{\mu}_t+s_1\tilde{e}^{\mu}_r+s_2\tilde{e}^{\mu}_{\ta}+s_3\tilde{e}^{\mu}_{\vp})\\
    \text{with }\;\Big(U=\frac{1}{\sqrt{1-s_1^2-s_2^2-s_3^2}}\Big).
\end{multline}
Here $s_1$, $s_2$, and $s_3$ are independent of each others but bounded by $-1$ and 1.
Using the SET expressions~\eqref{f6} and~\eqref{e6}, and the basis $\mathbf{\tilde{b}}$~\eqref{e4} along with the general rotating metric~\eqref{w1} (taking $K\equiv 1$), we find
\begin{equation}\label{we3}
I_{\text{WEC}}=U^2\times [\text{rhs of }~\eqref{n3}].
\end{equation}

We intend to apply~\eqref{we3} to conical timelike paths moving on the cone of equation $\ta =\text{constant}$ ($s_2\equiv 0$) from spatial infinity to the throat ($-1<s_1<0$) or conversely ($1>s_1>0$) and spiraling in the same or opposite direction as the rotating wormhole ($s_3>0$ or $s_3<0$, respectively, with $|s_3|<1$). For such paths, $u^{\mu}$ takes the form
\begin{equation}\label{we4}
u^{\mu}=U\Big(\frac{1}{N},\frac{s_1\sqrt{r-B}}{\sqrt{r}},0,\frac{aW}{N}+\frac{s_3}{r\sin\ta}\Big).
\end{equation}
Consider the rotating wormhole given by~\eqref{s3a}, \eqref{s4}, \eqref{ge3a}, and~\eqref{ge3b}. Equation~\eqref{we3} reduces to
\begin{align*}
&32\pi r^3N^2I_{\text{WEC}}/U^2=\{N [2N (B-r B')+4 r^2 N'' (r-B)\\
&-2N' (r B -2 r^2+r^2 B')]-3 a^2 r^4 W'^2 (r-B) \sin ^2\theta \}s_3^2\\
&+32a q^2 N W \sin  \theta  s_3+4rN^2B'-a^2 r^4 W'^2 (r-B) \sin ^2\theta\\
&+\{4N[2r(r-B)N'-BN]+a^2 r^4 W'^2 (r-B) \sin ^2\theta\}s_1^2.
\end{align*}
Fixing $n=6$, this reads
\begin{multline}\label{we9}
32\pi r^3I_{\text{WEC}}/U^2=(2A_3-3a^2A_2)s_3^2+2aCs_3\\+(32\pi r^3P_{r\,\text{s}}+a^2A_2)s_1^2+32\pi r^3E_{\text{s}}-a^2A_2,
\end{multline}
where $A_3=16\pi r^3P_{t\,\text{s}}$, $8\pi r^3P_{r\,\text{s}}$, and $8\pi r^3E_{\text{s}}$ are given by
\begin{align}\label{we10}
&A_3=r_0\frac{(4y-5)S_4+5}{y^4}\nn\\
&\quad +r_0\frac{[xy^3-(y-1)S_4-1](\sum_{i=1}^3\frac{S_i}{y^{i}})}{2y^4}>0,\\
&8\pi r^3P_{r\,\text{s}}=-r_0\frac{(y-1)S_4+1}{y^4}<0,\quad 8\pi r^3E_{\text{s}}=\frac{r_0x}{y}>0.\nn
\end{align}
In Ref.~\cite{types}, it was shown that $P_{t\,\text{s}}>0$ yielding $A_3>0$. In the slow rotation limit~\eqref{n6}, Eq.~\eqref{we10} simplifies to
\begin{equation}\label{we10a}
16\pi r^3I_{\text{WEC}}/U^2=A_3s_3^2+aCs_3+D_2,
\end{equation}
where
\begin{align}\label{we11}
D_2=&16\pi r^3(P_{r\,\text{s}}s_1^2+E_{\text{s}}),\nn\\
 \quad =&D_1+16\pi r^3\underbrace{P_{r\,\text{s}}(s_1^2-1)}_{>0}>D_1
\end{align}
The fact that $D_2>D_1$ shows that for any root $y_1$ to $D_1=0$ there corresponds a root $y_2<y_1$ to $D_2=0$. Since $A_3>0$ and $C>0$, we conclude that, for any given $s_1$ ($0<s_1^2<1$), the physical WEC is not violated along the conical timelike paths~\eqref{we4} for all possible values $x$ if 1) $y\geq y_2$ ($s_3$ arbitrary) or 2) $1\leq y< y_2$ and $-1<s_3\leq s_{3-}$ or $1>s_3\geq s_{3+}$ where ($s_{3-},s_{3+}$) are given by
\begin{equation}\label{we12}
s_{3\pm }=\pm \sqrt{\frac{-D_2}{A_3}}-\frac{aC}{2A_3}+\mathcal{O}(a^2)\qquad (1\leq y<y_2).
\end{equation}
If $y_2<1$ it is the case 1) that applies since $y\geq 1$. Note that if $1\leq y<y_2$ the expression inside the square root in~\eqref{we12} is smaller than 1 for all $0<s_1^2<1$. In case of no rotation, the conclusion remains valid with ($s_{3-},s_{3+}$) still given by~\eqref{we12} taking $a\equiv 0$.

In the above-made discussion, the timelike vector $u^{\mu}$ need not be geodesic. It is however straightforward to show that if the latter is geodesic, that is if $u^{\mu}_{\;;\nu}u^{\nu}=0$, the above-drawn conclusions remain valid. The geodesic equation $u^{\mu}_{\;;\nu}u^{\nu}=0$ with $u^{\mu}$ given by~\eqref{we4} is a set of four equations where $u^{\ta}_{\;;\nu}u^{\nu}$ is identically 0. Solving $u^{t}_{\;;\nu}u^{\nu}=0$ and $u^{\vp}_{\;;\nu}u^{\nu}=0$ we obtain
\begin{multline}\label{we13}
U=\frac{d_1-d_2aW}{N},\quad s_3=\frac{d_2N}{r(d_1-d_2aW)},\\ s_1^2=1-\frac{N^2}{(d_1-d_2aW)^2}\Big(1+\frac{d_2^2}{r^2}\Big),
\end{multline}
where ($d_1,d_2$) are real constants. The fact that $U\geq 1$ and ($N\to 1, W\to 0$) as $r\to \infty$ implies
\begin{equation}\label{we14}
    d_1\geq 1.
\end{equation}
The constant $d_2$ is not arbitrary, but rather subject to $-1<s_3<1$. The equation $u^{r}_{\;;\nu}u^{\nu}=0$ is automatically satisfied.

Now, restricting ourselves to the nonrotating case $a=0$, the rhs of~\eqref{we10a} becomes
\begin{equation}
\bar{A}_3s_3^2+\bar{D}_2
\end{equation}
with
\begin{multline}
\bar{A}_3=A_3+16\pi r^3(-P_{r\,\text{s}})>A_3,\\ \bar{D}_2=D_1+16\pi r^3(-P_{r\,\text{s}})\frac{N^2}{d_1^2}>D_1.
\end{multline}
Since, by~\eqref{s3a} and~\eqref{we14}, $N^2/d_1^2\leq 1$ and $\bar{A}_3>A_3$, the conclusions made in the paragraph following~\eqref{we11} and in the first paragraph following~\eqref{we12} remain valid on replacing $y_2$ by $\bar{y}_2$, which is a root to $\bar{D}2=0$.

\section{Conclusion \label{secc}}

We have shown that Teo's wormhole~\cite{teo} could be sourced by two anisotropic fluids. Applying the same procedure we constructed a redshift-free ($N\equiv 1$) rotating wormhole sourced by two anisotropic rotating fluids one of which is exotic and the other is a source-free electric or magnetic field. We have shown that the NEC is violated along any path crossing the throat in the nonrotating case but not in the rotating one.

Nonrotating wormhole with positive total energy density are classified into three types. Using a nonrotating massive type I wormhole~\cite{types} we constructed its rotating counterpart which both are sourced by two anisotropic fluids--exotic one and a source-free electromagnetic one. The shift and the shape functions of these rotating wormholes depend only on the mass while their angular velocity depends on the mass and the charge. We have shown the existence of a mass-charge constraint yielding almost no more dragging effects than ordinary stars.

We have proven the importance of nonrotating and rotating type I wormholes by showing the existence of conical spirals along which the physical NEC and WEC are not violated. Observes, particularly those crossing the throat, will not be able to measure negative amounts of energy densities in their frames if their journey borrows the conical spiral paths defined in this work and probably other paths too.

The analytical method developed in this work can easily be extended to account for the superposition of three fluids.




\end{document}